\documentstyle[12pt,epsf]{article}  
\newcommand{\be}{\begin{equation}}  
\newcommand{\ee}{\end{equation}}

\begin{document}  
\begin{titlepage}  
\pagestyle{empty}  
\noindent
\parbox{8cm}{\sf TSL/ISV-2000-237\\
                December 2000}
\vspace*{2cm}  
\begin{center}  
{  
\large\bf  
The photon structure and exclusive production of vector mesons
in $\gamma\gamma$ collisions
}  
\vspace{1.1cm}\\  
 {\sc L.~Motyka}$^{a,b}$,  
 {\sc B.~Ziaja}$^{c,a}$  
\vspace{0.3cm}\\  
  
$^a${\it High Energy Physics, Uppsala University, Uppsala, Sweden} \\   
$^b${\it Institute of Physics, Jagellonian University, Cracow, Poland}\\  
$^c${\it Department of Theoretical Physics, \\  
H.~Niewodnicza\'nski Institute of Nuclear Physics,  Cracow, Poland \\}

\end{center}  
\vspace{1.5cm}  

\begin{abstract}  
The process of exclusive vector meson production $\gamma\gamma \to 
J/\psi\,\rho^0$ is studied for almost real photons. This process may be 
reduced to photoproduction of $J/\psi$ off the $\rho^0$ meson. We 
discuss the  possibility of extracting the gluon distribution of $\rho^0$
and of the photon from such measurement. 
Predictions are also given for the reaction 
$e^+ e^- \to e^+ e^- J/\psi\, \rho^0$ for various $e^+e^-$ 
cms energies typical for LEP and for the future linear colliders.  
\end{abstract}

\end{titlepage}  
  
The successful operation of LEP2 has given a significant boost
for extensive studies of the photon structure --- both
theoretically and experimentally (see eg. \cite{EGSTRUCT,TGSTRUCT}).
We may expect further progress in this field after the next generation
of high energy linear colliders have started. However, up to now,
the existing results for the photon structure have rather limited
accuracy. They are based mainly on the inclusive data, i.e. 
the total $\gamma\gamma^*$ cross section. It should therefore be
useful to supplement this classical approach with some other 
independent methods of measurement of the partonic content of the photon.
In fact, the Ryskin process \cite{MGRPSI} of elastic vector meson 
photoproduction at HERA may serve as a good example of such 
alternative way of extracting gluon distribution of the proton 
\cite{RRML}. The cross section for the exclusive reaction 
$\gamma\gamma \to J/\psi V$ is expected to be dominated 
by light vector meson $V$ components of the photon. 
Thus, the exclusive reaction $\gamma\gamma \to J/\psi V$ should 
closely resemble the analogous well known one: $\gamma p \to J/\psi p$. 
Both of them have the advantage of directly testing the gluon
distribution with the enhanced sensitivity to the $x$~dependence at low $x$. 
Therefore  we shall analyze in detail the exclusive vector meson production 
$\gamma\gamma \to J/\psi V$ and show that, indeed, it may probe 
efficiently the gluon distribution of $V=\rho^0$, $\omega$ and $\phi$. 
Hence it may also constrain the gluon distribution of the photon.
This new proposed measurement, when combined 
with HERA data may also test the universality of the Regge behaviour
in exclusive vector meson production off different hadrons. 
Although such two photon reactions has been already studied,
in particular in Refs.~\cite{GINZ} and \cite{DONNA}, the connection
of the amplitude to the gluon distribution has never been
exploited before.     
Finally, in order to estimate the feasibility of this measurement we shall 
also compute the corresponding $e^+e^-$ cross sections.

In most of the successful models of the photon structure 
\cite{SAS,GRS,BKKS} it is assumed that 
the photon wave function (in the strongly interacting sector) is 
represented by two significantly different sets of configurations:
hadronic-like i.e. virtual light vector mesons~$V$
and perturbative ones, corresponding to $q\bar q$ 
or $q\bar q g$ systems with transverse momenta in the perturbative domain.
For the real photon the vector meson configurations dominate
the photon-proton cross section  (the Vector Meson Dominace model). 
The same property holds true 
for the $\gamma\gamma$ total cross section as well (see e.g. \cite{BKKS}). 
It is therefore legitimate to assume that in an exclusive process with a
light vector meson in the final state the dominance will be even
more pronounced. The perturbative configurations, less important
for the total cross section, will tend to produce hadronic
systems with higher invariant masses, not contributing
to the selected exclusive channel. Thus the exclusive process 
$\gamma \gamma \to J/\psi V$ occurs predominantly through 
vector meson component $V$ of one of the incoming photons.
It means that the scattering amplitude
${\cal M}(\gamma\gamma \to J/\psi V)$ will be well approximated 
by the amplitude  ${\cal M}(\gamma V \to J/\psi V)$ multiplied
by the photon-meson coupling $\sqrt{\alpha}g_V$.

It was pointed out by Ryskin \cite{MGRPSI} that 
the elastic $J/\psi$ photoproduction off the proton at 
the $\gamma p$ cms energy $W$ may probe
the gluon distribution of the proton $xG^p(x,Q^2)$
for $x=M_{J/\psi}^2/W^2$ and $Q^2=M_{J/\psi} ^2 /4$. Namely,
the obtained differential cross section for $t=0$
\be
{d\sigma_{\gamma p\to J/\psi p} \over dt} (W^2,t=0) = 
{16\pi^3 [\alpha_s(M_{J/\psi}^2/4)]^2 \Gamma_{ee} ^{J/\psi} 
\over 3\alpha M^5 _{J/\psi}}
\left[ xG^p(x,M_{J/\psi}^2/4) \right]^2
\label{dsdt1}
\ee
depends on the square of $xG^p (x,M_{J/\psi}^2/4)$ which allows
for the direct measurement of this quantity with enhanced
sensitivity. This concept has proven to be extremely succesful, and it
strongly stimulated the experimental efforts at HERA \cite{HERAVM,HERALVM}. 
Since the first paper \cite{MGRPSI} significant
theoretical progress in this subject has also been made \cite{NEWERPSI,MRT}. 
A set of important corrections to the leading result (\ref{dsdt1})
was isolated and studied in detail. We shall present them, 
following the discussion in Refs~\cite{RRML} and \cite{MRTUPS}.
In particular, it was shown that relativistic corrections 
may be absorbed into the constituent mass of the charmed quark, 
$m_c = M_{J/\psi} /2$ \cite{RRML, RELCORR}. 
Formula (\ref{dsdt1}) takes into account only the dominant imaginary 
part of the complex amplitude ${\cal M}(\gamma p \to J/\psi p)$. 
The estimate of the smaller real part of ${\cal M}(\gamma p \to J/\psi p)$
may be obtained in a standard way from the Regge model. 
The additional enhancement due to the real part of the matrix
element may be described  by a multiplicative correction factor:
\be
C_{rp} \simeq 1+{\pi^2 \lambda ^2 \over 4}
\ee
with $1+\lambda$ defined as the (local) pomeron intercept:
\be
\lambda = {\partial \log [xG(x,Q^2)] \over \partial \log (1/x)}. 
\ee  

The QCD NLO corrections to the $\gamma J/\psi$ impact-factor \cite{RRML} 
introduce an additional factor of 
\be
C_{NLO}\simeq 1 + {\alpha_s (M_{J/\psi}^2 /4) \over 2}.
\ee  
It has been pointed out that there are non-negligible corrections
coming from the fact that the parton distribution entering
Eq.~(\ref{dsdt1}) should be in fact off-diagonal \cite{OFFDIAG}.
The detailed study \cite{MROFFDIAG} showed that in order to 
account for this effect the cross section should be multiplied by 
a factor 
\be
C_{off} \simeq 1.2.
\ee
The value of $C_{off}$ is governed mainly by the degree to which 
the parton distribution is off-diagonal in the production process 
and should not be too sensitive to differences in the structure of 
$\rho^0$ and the proton and to the  choice of~$x$.
It seems that the effects of rescattering of the $c\bar{c}$ pair 
and of the exchanged gluon transverse momenta partially cancel each
other \cite{RRML}. The overall effect is small hence
including these corrections is not neccessary to maintain 
compatibility between the theoretical and the experimental results
\cite{MRTUPS}.

For the sake of lowering statistical errors it is  favourable to
study the total cross section $\sigma_{\gamma p\to J/\psi p}(W^2)$.
Due to the (approximate) exponential decrease of $d\sigma / dt$:
\be
{d\sigma_{\gamma p\to J/\psi p} \over dt}  (W^2,t) 
\simeq 
{d\sigma_{\gamma p\to J/\psi p} \over dt} (W^2,t=0) 
\exp (B_{J/\psi p} t)
\label{rystot}
\ee
with $B_{J/\psi p}$ equal to $5~{\rm GeV}^{-2}$ and very weakly depending 
on $W^2$ \cite{HERAVM}, one may write
\be
\sigma_{\gamma p\to J/\psi p}(W^2) = {1 \over B_{J/\psi p}}
{d\sigma_{\gamma p\to J/\psi p}  \over dt}(W^2,t=0).
\ee
Recent refined studies \cite{RRML,MRT,MRTUPS} have shown 
that the theoretical results agree very well with the data from HERA 
\cite{HERAVM}. 
Of course, it is straightforward to generalize formula (\ref{dsdt1}) 
for the case of production of $J/\psi$ off $\rho^0$, $\omega$ and $\phi$. 
There are two different approaches to model the structure of 
vector mesons. The first one is based on the expected, approximate
similarity between spatial wave functions of pions and light vector mesons
which implies $xG^V(x,Q^2) \simeq xG^\pi(x,Q^2)$ \cite{GRS}.
Another point of view is presented in Ref.~\cite{SAS}, in  which 
the differences between vector mesons and pions (e.g. spin, lifetime) are 
underlined and the parton distributions of $V =\rho^0$, $\omega$ and $\phi$ 
are fitted to describe the experimental results for the photon structure.
We shall include in our analysis both possibilities.

Certainly, there arise subtleties. First, there exist no direct 
experimental results for $xG^V (x,Q^2)$
and the existing parametrisations are based on models.
So the results we shall obtain will be uncertain.
However, this analysis is more aimed to estimate the feasibility
of the measurement than to provide a precise answer. 
Besides, it makes the proposed measurement the first one
to probe directly gluonic content of light vector mesons.
Second, the value of $B_{J/\psi\,\rho}$ for the exclusive
production off the light vector mesons has never been measured.
Fortunately, it may be constrained by theoretical considerations
on the basis of indirect data. With a reasonable accuracy 
the $B$ coefficients in the studied processes ($pp \to pp$, 
$\gamma p \to V p$, $\gamma^* p \to V p$ and $\gamma p \to J/\psi p$) 
may be described as additive combinations of  characteristic components
$b_i$ for the hadrons $h_i$ taking part in the process i.e.
$B_{ij} = b_i + b_j$.
In the dipole picture of the diffractive scattering these components
may be related to the mean squared sizes of the contributing dipoles
(see e.g. \cite{NIKOLAEV}). For energies 50~GeV$< W <$100~GeV
the following values of $B$ has been reported:
for elastic $pp$ scattering  $B_{pp}\simeq 10-12~{\rm GeV}^{-2}$,  
for $\rho^0$ photoproduction at HERA $B_{\rho p} \simeq 11~{\rm GeV}^{-2}$,
and it decreases with $Q^2$ when the photon is virtual \cite{HERALVM}, and
finally for the $J/\psi$ photo- and electroproduction off protons
$B_{J/\psi p} \simeq 5~{\rm GeV}^{-2}$. Taking into acount that
$ b_{J/\psi} \ll b_p$, we arrive at the conclusion that
$b_p \simeq b_\rho \simeq B_{J/\psi \rho} \simeq  5.5~{\rm GeV}^{-2}$.
Due to the experimental errors for $B$ coefficients 
(typically $\pm$1~GeV$^{-2}$) and 
the crudeness of the estimation of the latter result 
we adopt $B_{J/\psi \rho} \simeq  (5.5 \pm 1.0)~{\rm GeV}^{-2}$.
Finally, the corrections due to rescattering may differ between
the production off the proton and off the light vector mesons.
This difference should not be significant for the result because the 
correction is small anyway, moreover, as suggested by the 
relation  $b_p \simeq b_\rho$ the sizes of the proton and $\rho^0$ are 
similar.  
So, for the estimate of the differential $\gamma\gamma \to J/\psi\,\rho^0$
cross section we shall use the following formula:
\[
{d\sigma_{\gamma\gamma \to J/\psi\, \rho} \over dt}(W^2,t) =
 C_{off} C_{NLO} C_{rp} \alpha g_{\rho}^2 \times
\]
\be
{16\pi^3 [\alpha_s(M_{J/\psi}^2/4)]^2 \Gamma_{ee} ^{J/\psi} 
\over 3\alpha M^5 _{J/\psi}} \left[ xG^p(x,M_{J/\psi} ^2/4) \right]^2
\exp(B_{J/\psi\,\rho} t),
\label{master}
\ee 
where $W$ denotes  $\gamma\gamma$ cms energy, $x= M_{J/\psi}^2/W^2$ and 
$g_{\rho} ^2 = 0.454$.

The real opportunity to study high energy photon-photon collisions
is provided so far only in $e^+ e^-$ colliders mainly at LEP2 
\footnote{It might be also possible to test directly such photon-photon 
processes at future photon colliders.}.
There, a large fraction of events may be described in terms of
a factorizing $\gamma\gamma$ cross section and known photon fluxes 
in the colliding leptons \cite{EGSTRUCT}:
\be
d\sigma_{e^+ e^- \to e^+ e^- X}  = 
f_{\gamma/e}(z_1,Q_1 ^2) f_{\gamma/e}(z_2, Q_2 ^2)
   \sigma_{\gamma\gamma \to X} (W^2) dQ_1 ^2\, dQ_2 ^2\, dz_1 \, dz_2,
\label{sigee1}
\ee
where $W^2 = z_1 z_2 s_{ee}$, $s_{ee}$ is 
the cms leptonic collision energy squared,
and the flux factor for the photon carrying the fraction 
$z$ of the incoming electron and the virtuality $Q^2$ reads:
\be
f_{\gamma/e}(z,Q^2) = {\alpha \over 2\pi}
\left[ {1\over Q^2} {1+(1-z)^2 \over z} - {2 z m_e ^2 \over Q^4} 
\right], 
\ee
where $m_e$ is the electron mass. The lower limit for $Q_i ^2$ follows 
directly from kinematics of the process:
\be 
Q_{min} ^2 = {m_e^2 z^2 \over {1-z}}.
\ee
In the process $\gamma(Q_1 ^2) \gamma(Q_2 ^2) \to J/\psi\,\rho^0$ 
(with $J/\psi$ moving in the direction of $\gamma(Q_1 ^2)$)
the cross section $\sigma_{\gamma(Q_1 ^2) \gamma(Q_2 ^2) \to J/\psi\,\rho}$
depends only weakly on $Q_1 ^2$ ($Q_2 ^2$) when $Q_1 ^2< M_{J/\psi} ^2$
($Q_2 ^2 < M_{\rho} ^2$). On the other hand, the cross section falls down
rapidly for large $Q_i^2$.  
This follows from the fact, that in the processes of this type 
the cross section is rather a function of $M_V^2+Q^2$ than 
of $Q^2$ alone.  
So, for untagged measurements we may account for this property of 
$\sigma_{\gamma(Q_1 ^2) \gamma(Q_2 ^2) \to J/\psi\,\rho}$
by the standard procedure of taking 
$\sigma_{\gamma(Q_1 ^2) \gamma(Q_2 ^2) \to J/\psi\,\rho}$
for $Q_1 ^2=Q_2 ^2=0$ and imposing the limits $Q_1 ^2< M_{J/\psi} ^2$
and $Q_2 ^2 < M_{\rho} ^2$. The dependence of the $e^+ e^-$ 
cross section integrated over $Q_i^2$ on the choice of the upper limits
$Q_{i,max} ^2$ is only logarithmic, so there is no need for a more 
detailed treatment. 
With this approximation we may integrate over $Q_i^2$ to obtain:
\be
 d\sigma_{e^+ e^-}  = 
f^{(1)}_{\gamma/e}(z_1) f^{(2)}_{\gamma/e}(z_2)
   \left. \sigma_{\gamma\gamma \to J/\psi\,\rho} (W^2) 
    \right|_{Q_1^2=Q_2^2 =0}
    dz_1 \, dz_2
\label{sigee2}
\ee
with
\be
f^{(i)}_{\gamma/e}(z) = {\alpha \over 2\pi}
\left[  {1+(1-z)^2 \over z} 
\log\left( Q^2 _{i,max} \over Q^2 _{min} \right) 
  - 2 z m_e ^2 \left( {1 \over  Q^2 _{min}} - {1 \over  Q^2 _{i,max}} \right)
\right],
\ee
where $ Q^2 _{1,max} = M^2_{J/\psi}$ and $ Q^2 _{2,max} = M^2 _{\rho}$.

Let us also introduce a convenient variable to define corresponding
differential cross section 
\be
Y = \log {W^2 \over M^2 _{J/\psi}},
\ee
i.e. $Y$ corresponds to the  $\log (1/x)$. 
It then is useful to define:
\be
{d\sigma_{e^+e^-} \over dY}(Y) = 
\int {d\sigma_{e^+ e^-} \over dz_1 \, dz_2} 
\delta\left(Y-\log(z_1 z_2 s_{ee} / M^2 _{J/\psi}) \right)\,
dz_1\, dz_2.
\label{dsdy}
\ee
In fact, in the preceding analysis we assumed that $J/\psi$ follows 
the direction of a particular photon, and if we intend to describe the 
process in the inclusive way i.e. irrespectively of the direction of the
momentum of the  produced mesons, our results should be multiplied by 
a factor~2. Whenever we use this ``direction inclusive'' cross section, 
 we shall state it explicitly.
 
\begin{figure}[hbpt] 
\begin{center} 
\leavevmode 
\epsfxsize = 11cm 
\epsfysize = 8cm 
\epsfbox{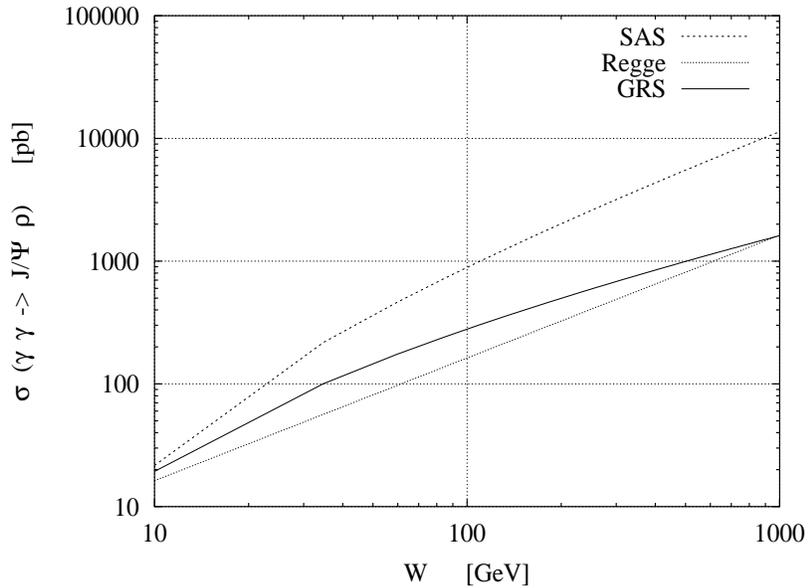} \\ 
\end{center} 
\caption{ \small Energy dependence of the cross section 
$\sigma _{\gamma \gamma \to J/\psi\,\rho}(W)$ 
for various parametrisations of the vector meson gluonic distribution.
Solid line corresponds to the calculations with GRS-NLO parametrisation of the
gluon distribution, dashed line represents the results obtained for 
SaS1D parametrisation of $xG^V(x,Q^2)$, and dotted line shows the cross 
section for Regge-motivated (Eq.~(\ref{regge})) gluon distribution.}
\label{siggg}
\end{figure}

In Fig.~\ref{siggg} we display the set of predicted photon-photon
cross sections $\sigma_{\gamma\gamma \to  J/\psi\,\rho} (W^2)$
obtained with different assumptions concerning the gluon distribution 
of $\rho^0$. Thus we plot the results for the GRS-NLO 
parametrisation \cite{GRS} and SaS1D parametrisation \cite{SAS}. 
We also include a phenomenological Regge motivated 
ansatz for $xG^V(x,M_{J/\psi}^2 /4)$  by setting for $x<x_0$
\be
xG^V(x,M_{J/\psi}^2 /4) = x_0 G^V(x_0,M_{J/\psi}^2 /4) 
\left({x_0 \over x} \right)^{\lambda_P}
\label{regge}
\ee
with $x_0 = 0.1$ and $\lambda_P = 0.25$ in accordance with HERA data 
(see also Ref.~\cite{DONNA}).
To obtain the value of $x_0 G^V(x_0,M_{J/\psi}^2 /4)$ the GRS-NLO  
parametrisation is used. 
Note, that because of the different $x$-dependencies i.e. the different
values of $\lambda$ the correction factors $C_{rp}$ accounting for the 
real part of the amplitude are different for the three curves.
For the rather small $\gamma\gamma$ energy $W=20$~GeV, our results 
for the cross section are larger by a factor between 1.5 (Regge motivated) 
and 4 (SaS1D) than those obtained in \cite{DONNA}. 
The steepness of the rise with increasing energy~$W$ is strongly 
dependent upon the choice of the gluon parametrisation.
The cross sections presented in Fig.~\ref{siggg} might be directly
probed at future photon colliders providing constraints
on the vector meson structure.

\begin{figure}[hbpt] 
\begin{center} 
\begin{flushleft} {\large\bf a)}\\ \end{flushleft}
\leavevmode 
\epsfxsize = 11.0cm 
\epsfysize = 7.5cm 
\epsfbox{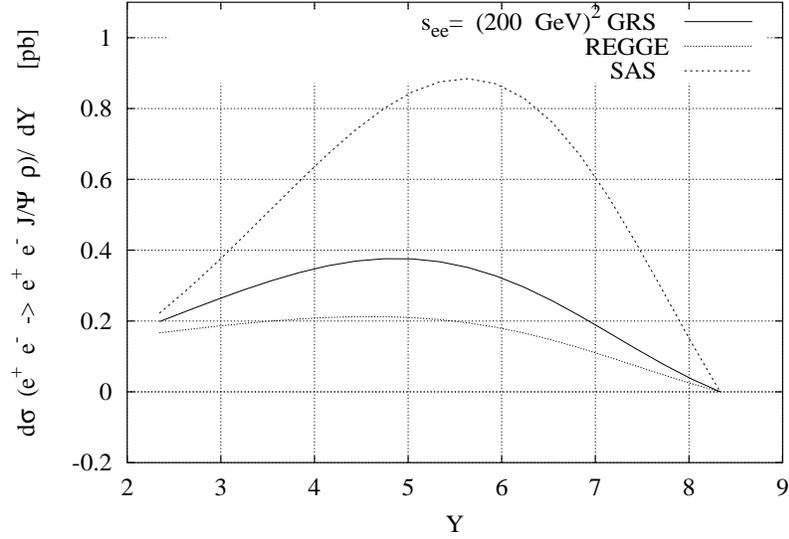} \\ 
\begin{flushleft} {\large\bf b)}\\ \end{flushleft}
\leavevmode 
\epsfxsize = 11.0cm 
\epsfysize = 7.5cm 
\epsfbox{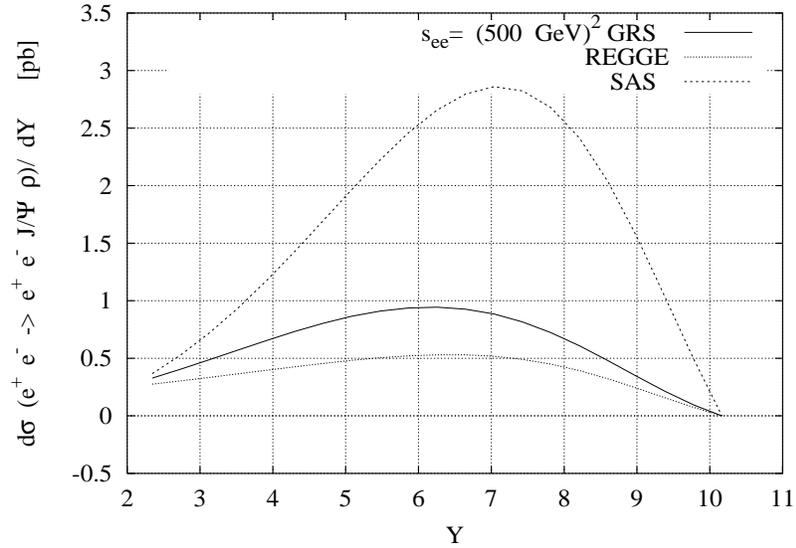} \\ 

\end{center} 
\caption{ \small Cross section $d\sigma_{e^+e^-} /dY (Y)$ (Eq.~(\ref{dsdy})) 
plotted for three different choices of the  gluon distribution and 
for two $e^+ e^-$ collision energies $\sqrt{s_{ee}}$ = 200 (a) and 500~GeV (b).
Solid line corresponds to the calculations with GRS-NLO parametrisation of the
gluon distribution, dashed line represents the results obtained for 
SaS1D parametrisation of $xG^V(x,Q^2)$, and  dotted line shows 
the cross section 
for Regge-motivated (Eq.~(\ref{regge})) gluon distribution.}
\label{sigeeY}
\end{figure}

In Fig.~\ref{sigeeY} we show $d\sigma_{e^+e^-} / dY $ as defined 
by formula (\ref{dsdy}) for two $e^+e^-$ cms energies:
$\sqrt{s_{ee}} = $200 (a) and 500~GeV (b) respectively, and
for GRS-NLO, Regge-motivated and SaS1D choices of the gluon distribution 
of the vector meson. The energies 
are chosen to match LEP2 and future linear collider conditions.
The differences in the shape and normalisation of the curves in 
Fig.\ ~\ref{sigeeY} are large. Even though the normalisation factors 
(e.g. the uncertainty of $B_{J/\psi\rho}$) introduce
some uncertainty into formula (\ref{master}), it still remains 
possible to distinguish between the models.

In Tab.~1 the total (``direction inclusive'')  $e^+e^-$ cross sections for
$J/\psi\,\rho^0$ production, with a cut $W>W_0 = 10$~GeV (introduced 
in order to stay in the diffractive regime), are listed for all 
the three considered models of the gluon distributions and for the energies  
$\sqrt{s_{ee}}$ = 90, 200, 500 and 1000~GeV. In particular, 
we may read out from the table that at LEP2 with an integrated
luminosity of about 700~pb$^{-1}$ per experiment we may expect to 
have a sizable amount of 1300 to 4900 events.
In future colliders, like TESLA \cite{TESLA}, the integrated 
luminosity per year may reach 50~fb$^{-1}$ which would correspond to 
0.3 to 1.4 million of events. Certainly, the number of interesting events 
which can be  uniquely identified will be severly cut down when the 
acceptance is taken into account, mainly  because  the $J/\psi$
may be reliably measured only through its leptonic decay products.
Another experimental difficulty may arise because, as one may see in
the HERA data \cite{HERAVM}, the total cross section for 
the proton dissociative  photoproduction of $J/\psi$
is larger than for the elastic process. This should also hold
for the production of $\rho^0$. Besides, there may also 
occur a hard diffraction of one photon accompanied by 
$\gamma \to J/\psi$  transition on the other side.
Such processes produce background for the  exclusive
processes that we want to focus on, 
however, their impact on the feasibility of the 
measurement depends strongly on particular experimental conditions.
For instance, at HERA, the different classes of events 
(elastic and proton dissociative) are identified with 
high efficiency. 
Of course, the given results may be easily generalized to the
other neutral vector meson species both within the heavy 
(e.g. $\psi'$, $\Upsilon$) and light ($\omega$, $\phi$) sectors
correspondingly. 

It can also be seen that the measurement proposed by us
may be useful to distinguish between the models of the gluon distribution 
of light vector mesons i.e. the non-perturbative component of the
photonic gluon distribution. In principle, the region of low $x$ would be 
probed at LEP2 ($x>10^{-3}$) and future linear colliders ($x>10^{-4}$).
Both the shape of the differential 
cross section $d\sigma_{e^+e^-} / dY$ (see Fig.~\ref{sigeeY}) and the value of 
the total cross section depend significantly on the model. 
On top of that there is an interesting, model independent question of how 
the $J/\psi$ photoproduction off $\rho^0$ differs from the photoproduction 
off proton, in particular, if values of the pomeron intercepts 
$1+\lambda_P$ characterizing these processes are the same.

\begin{table}
\begin{center}
\begin{tabular}{|c|ccc|}
\hline
$\sqrt{s_{ee}}$ [GeV] & \multicolumn{3}{|c|}{$\sigma(e^+e^- \to e^+e^- J/\psi\,\rho^0)$ [pb]} \\
\cline{2-4}
    &  GRS   & SaS1D  & Regge \\
\hline
90   &  0.9  &  1.7  & 0.6   \\
200  &  3.1  &  7.0  & 1.9   \\
500  &  9.8  & 27.2  & 6.0   \\  
1000 &  21 & 68  & 13  \\ 
\hline
\end{tabular}
\end{center}
\caption{\small
The total cross section for the process 
$e^+e^- \to e^+e^- J/\psi\,\rho^0$ for the 
$\gamma\gamma$ energy $W>10$~GeV. The values are displayed for four 
$e^+e^-$ collision energies typical for the existing and future accelerators 
and for three different parametrisations of the gluon distribution. 
}
\label{tab1}
\end{table}

To conclude, we studied in detail the exclusive process
$\gamma\gamma \to J/\psi\,\rho^0$ using as crucial ingredients 
the Vector Meson Dominance model and the Ryskin picture of 
elastic vector meson production off a hadron. Measurement of this process
could provide for the first time direct information of the gluon 
distribution in light vector mesons. 
It would also constrain parametrisations of the photonic gluon distribution
in the low~$x$ domain. 
It may be possible to perform a successful experimental analysis 
already using LEP2 data, and at future linear colliders the  
number of events would be large enough to perform a thorough study.

\section*{Acknowledgements}
We are very grateful to J.~Kwieci\'{n}ski for enlightening discussions
at all stages of this research. We thank  K.~Doroba for 
informing us about the LEP signal of $J/\psi$ diffractive production,
B.~Bade\l{}ek  and G.~Ingelman for critically reading the 
manuscript and very useful comments.
L.\ M.\ is grateful to the Swedish Natural Science Research Council 
for the postdoctoral fellowship.
B.\ Z.\ is a fellow of the Stefan Batory Foundation.
This research has been supported in part by the EU Fourth Framework 
Programme `Training and Mobility of Researchers',  
Network `Quantum Chromodynamics and the Deep Structure of Elementary  
Particles', contract FMRX--CT98--0194 and by the Polish Committee for 
Scientific Research (KBN) grants Nos 2~P03B~04718 and 2~P03B~05119.

\end{document}